\documentclass[twocolumn,prl,showpacs,superscriptaddress]{revtex4} 
\usepackage{t1enc,graphicx,amsmath,amstext,amsfonts,amssymb,amsxtra,upgreek,color,hyperref}
\usepackage{graphicx}
\usepackage{bm}
\usepackage[latin1]{inputenc}
\usepackage{xspace}
\usepackage{latexsym}
\usepackage{pst-plot}
\usepackage{pstricks-add}

 \newcommand{\bs}{\boldsymbol}

\newcommand{\kB}{k_{\text{B}}}  
\newcommand{\microns}{\ensuremath{\upmu\text{m}}}

\newcommand{\ascat}{\ensuremath{a_{\text{3d}}}}
\newcommand{\OD}{\text{OD}}

\begin{document}

\title{The equilibrium state of a trapped two-dimensional Bose gas}
\author{Steffen\,P. Rath}\thanks{Now at Physik Department, TU M\"unchen, Germany.} 
\author{Tarik Yefsah}
\author{Kenneth\, J. G\"unter}
\author{Marc Cheneau}\thanks{Now at MPI für Quantenoptik, Garching, Germany.}
\author{R\'emi~Desbuquois} 
\affiliation{Laboratoire Kastler Brossel, CNRS, UPMC, \'Ecole Normale Sup\'erieure, 24 rue Lhomond, 75005 Paris, France}
\author{Markus Holzmann} 
\affiliation{LPTMC, CNRS, UPMC,  4 Place Jussieu, 75005 Paris and LPMMC, CNRS-UJF, BP 166, 38042 Grenoble, France} 
\author{Werner Krauth} 
\affiliation{Laboratoire de Physique Statistique, CNRS, UPMC, Universit\'e Paris Diderot, \'Ecole Normale Sup\'erieure,  24 rue Lhomond, 75005 Paris, France}
\author{Jean Dalibard} 
\affiliation{Laboratoire Kastler Brossel, CNRS, UPMC, \'Ecole Normale Sup\'erieure, 24 rue Lhomond, 75005 Paris, France}

\date{\today}

\begin{abstract} 
We study experimentally and numerically the equilibrium density profiles of a trapped two-dimensional $^{87}$Rb Bose gas, and investigate the equation of state of the homogeneous system using the local density approximation.  We find a clear discrepancy between in-situ measurements and Quantum Monte Carlo simulations, which we attribute to a non-linear variation of the optical density of the atomic cloud with its spatial density. However, good agreement between experiment and theory is recovered for the density profiles measured after time-of-flight, taking advantage of their self-similarity in a two-dimensional expansion.
\end{abstract}

\maketitle

Low-dimensional atomic gases provide stringent tests of the many-body description of quantum matter, because thermal or quantum fluctuations play a more important role than for three-dimensional (3d)  fluids \cite{Mermin:1966}. These systems are prepared by freezing one or two motional degrees of freedom \cite{Gorlitz:2001b}. For 2d Bose gases, recent experiments  \cite{Hadzibabic:2006,Clade:2009} and corresponding numerical analyses \cite{Holzmann:2008a,Bisset:2009b} gave evidence for a Berezinskii--Kosterlitz--Thouless transition, with a quasi-long range order of the phase of the gas below a critical temperature. 
 
A remarkable feature of the uniform 2d Bose gas is the scale invariance of its equation of state.  For a large domain of parameters, the phase space density $D=n\lambda^2$ is not an independent function of the chemical potential $\mu$ and the temperature $T$, but depends only on the ratio $\alpha=\mu/\kB T$. Here $n$ is the 2d spatial density and $\lambda=(2\pi\hbar^2/m\kB T)^{1/2}$ is the thermal wavelength. Scale invariance originates from the fact that the interaction strength in 2d is determined by a dimensionless number that is approximately energy-independent, $\tilde g=\sqrt{8\pi}\,\ascat/\ell_z$, where $\ascat$ is the scattering length characterizing low energy interactions in 3d, and $\ell_z$ is the thickness of the gas along the frozen direction $z$ [for harmonic confinement with frequency $\omega_z$, $\ell_z=(\hbar/m\omega_z)^{1/2}$] \cite{Petrov:2000a}. Since interactions provide no energy scale, dimensional analysis implies that $D$ can be cast in the form $D=F(\alpha,\tilde g)$. This holds when $\ell_z\gg \ascat$, and applies for all experiments so far.

In this Letter we present results from combined experimental and numerical studies to test this scaling property of the 2d Bose gas. The experiments are performed with $^{87}$Rb atoms and the results are compared with Quantum Monte Carlo (QMC) simulations. Experimental density distributions are inferred from the absorption of a resonant probe light beam. We investigate both the in-situ distribution of the gas and the one obtained after a time-of-flight (TOF) in the $xy$ plane. The numerical results confirm the scale invariance and are in excellent agreement with the prediction based on the equation of state for the uniform 2d gas \cite{Prokofev:2002} and on the local density approximation. 
The measured in-situ distributions clearly differ from the numerical predictions if we assume a linear relation between the optical density of the cloud and its spatial density $n$. However we point out that the usual single-scattering approximation for the probe beam photons, which is at the basis of this linear relation, is insufficient in our situation. Indeed the inter-particle distance in the center of the trap is comparable to $k_{\rm L}^{-1}$, where $k_{\rm L}$ is  the wave vector of the probe light.  After long TOF durations the densities are considerably lower, and the single scattering approximation holds. We then recover good agreement between theory and experiment, using a second (dynamical) scale invariance of the 2d Bose gas confined in a harmonic potential \cite{Pitaevskii:1997a}.

\begin{figure}[t] 
\begin{center}
\includegraphics[width=80mm]{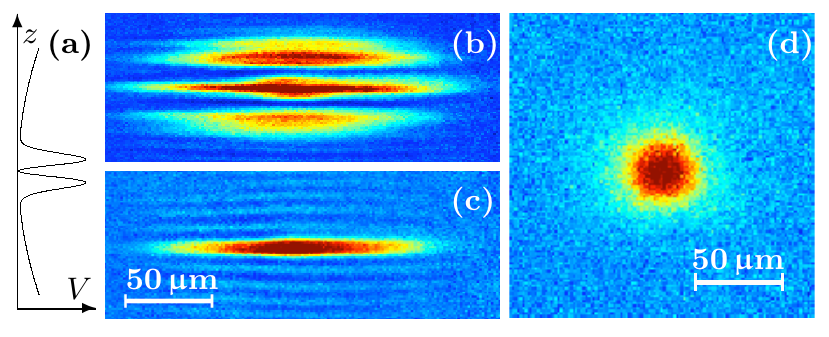} 
\end{center}   
\vskip -3mm
\caption{(Color online) (a) Potential $V$ along the vertical direction $z$ created by the magnetic trap and the laser beam.     (b-c) Side view of the cloud before (b) and after  (c) depumping atoms in the side wells. The horizontal stripes are due to  diffraction. (d) Top view in-situ image yielding fit parameters $T=132\,$nK and $\alpha=0.29$ (for $\xi=0.25$). } \label{fig:cleaning} 
\end{figure}

In our experiment, we first prepare a Bose-condensed gas of $\sim 3\times 10^5$ $^{87}$Rb atoms in the $F=2,\;m_F=2$ hyperfine state of the ground level $5^2$S$_{1/2}$. The gas is confined in a magnetic Time-averaged Orbiting Potential trap \cite{Petrich:1995} at a temperature of 160\,nK, obtained through evaporative cooling with a radio-frequency (rf) field. We then add a dipole potential providing strong confinement in the vertical direction. This potential is generated by a laser beam at a wavelength of 532\,nm and a power of 1.3\,W. The beam passes a holographic plate   that imprints a phase of $\pi$ on its upper half, and is focussed onto the atoms \cite{Smith:2005}. There it creates (together with the magnetic trap) the potential illustrated in Fig.~\ref{fig:cleaning}a. The vertical and horizontal waists of the beam in the absence of the holographic plate are 5.0\,\microns\ and 140\,\microns, respectively. The trapping frequencies in the combined potential, measured by exciting the dipole oscillation of the atom cloud, are $\omega_z/2\pi=3.6\,(3)$\,kHz along the vertical axis, leading to $\tilde g=0.146\,(6)$, and $\omega_x/2\pi= 21.0\,(5)\,\text{Hz}$ and $\omega_y/2\pi=18.8\,(5)\,\text{Hz}$ in the horizontal plane. For convenience we define $\omega=(\omega_x \omega_y)^{1/2}=2\pi\times 19.9$\,Hz. 

The dipole potential is ramped up in 1.5\,s and it cuts out a plane of atoms in the center of the original cloud, while the remaining atoms reside in the side wells (Fig.~\ref{fig:cleaning}b). We depump these unwanted atoms to a non-detected state ($F=1$) with a 35\,ms laser pulse tuned to the $F=2 \leftrightarrow  F'=2$ transition of the $D_2$ line, masking the central plane of atoms (Fig.~\ref{fig:cleaning}c). The atoms are held in the combined trap during $5$\,s in the presence of the rf which controls the final temperature. 

We probe the spatial density distribution in the $xy$ plane using absorption imaging of a resonant probe laser beam propagating along the vertical axis.  The measurement provides a map of the optical density ($\OD$) of the atomic cloud, defined as the natural logarithm of the ratio between incident and transmitted light intensities. Neglecting multiple scattering of photons, one has $\OD(\bs r)=\sigma n(\bs r)$, where $\sigma$ is the absorption cross-section. Absorption imaging can be performed in the presence of the trap magnetic field (in-situ measurement), or after a TOF expansion in the $xy$ plane.  A typical in-situ image is shown in Fig.~\ref{fig:cleaning}d. From each image we generate a radial density profile $\OD(r)$ by averaging over the azimuthal angle while accounting for the residual ellipticity in the $xy$ plane. 

We determine $\mu$ and $T$ by fitting the profile $\OD(r)$ with the mean-field Hartree-Fock (MFHF) prediction, which reads for a strictly 2d gas  (see e.g.~\cite{Hadzibabic:2009red}): $D=-\ln \left( 1-e^{\alpha -\tilde g D/\pi} \right)$. Since our highest temperatures ($\sim 150$\,nK) are comparable to $\hbar \omega_z/\kB=170$\,nK, the gas is only quasi-2d  \cite{Holzmann:2008,Hadzibabic:2008,Bisset:2009a} and we take into account corrections to $D(\alpha)$ due to the residual excitation of the atom motion in the $z$ direction \cite{Holzmann:2008a,Holzmann:2008,Hadzibabic:2008}. We restrict the fitting domain to simultaneously fulfill two conditions: (i) the phase space density in the axial ground state must be lower than 2.5, so that beyond-mean-field corrections are negligible \cite{Holzmann:2008,Lim:2009}; (ii)  the optical density of the cloud must be lower than 0.2 to exclude distortions due to multiple scattering of probe photons (see below). Because the spatial density enters non-linearly in the relation between $D$ and $\alpha$, the fitting procedure can also provide a value of the detection efficiency $\xi$, defined as the ratio between the actual absorption cross-section and the ideal one expected for monochromatic probe light in the absence of stray fields. For all images corresponding to a given TOF duration we extract a single value of $\xi$ from the fit. For in-situ images we find the small value $\xi=0.25\,(4)$, which accounts for the strong reduction of  the absorption cross-section due to the magnetic field of the trap. With this correction factor, an optical density of $0.1$ corresponds to a density $n\simeq 3.0\,\microns^{-2}$. 

Our path-integral QMC simulations are performed in the canonical ensemble in 3d continuum space, for the same geometry as the experiments. They take into account residual excitations along the strongly confined direction $z$ \cite{Holzmann:2008a,Holzmann:2009red}. Pair interactions are described by a 3d pseudo-potential \cite{Krauth:1996}. All thermodynamic properties of the gas are obtained to high precision and without systematic errors. The chemical potential associated to a given atomic distribution is obtained from a fit of the wings of the distribution with the MFHF prediction, as for experimental data.

\begin{figure}[tbp]
\begin{center}
\includegraphics[width=80mm]{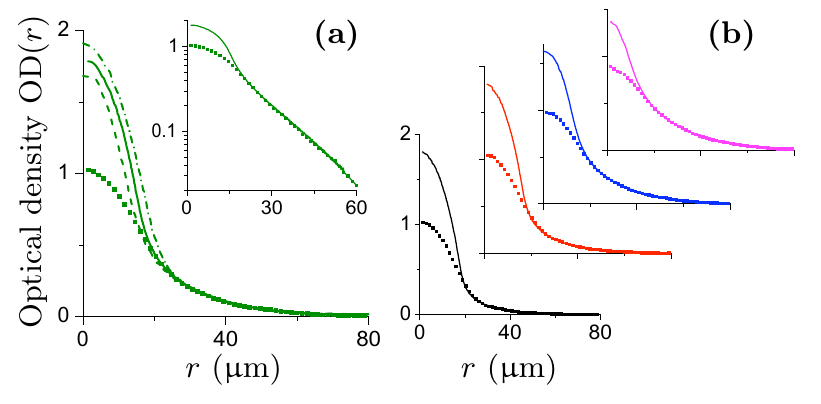}
\end{center}   
\vskip -3mm
\caption{(Color online) (a) Dots: Measured in-situ profile $\OD(r)$. For $\xi=0.25$ the fit with MFHF theory yields $T=126\,(6)\,$nK, $\alpha=0.34\,(9)$, where uncertainties represent standard deviations obtained by fitting individual images. Continuous line: corresponding QMC simulation with  $N=73900$ atoms (inset: same data in log plot). Upper (dash-dotted) and lower (dashed) lines: QMC results obtained assuming $\xi=0.21$ [fit parameters  ($T\;{\rm(nK)},\,\alpha,\, N$)=(130,\,0.39,\,96300)] and  $\xi=0.29$ (122,\,0.29,\,57900), respectively. (b) Set of measured density profiles (dots) and corresponding QMC results (lines) for rf evaporation parameters. From bottom to top: ($T\;{\rm(nK)},\alpha,N$)=(87,\,0.49,\,54100) [black], (109,\,0.39,\,63800) [red], (142,\, 0.28,\,78400) [blue], (153,\, 0.23,\,79900) [magenta]. Each experimental profile in (a) and (b) is an average of 9 images.}
  \label{fig:insitu}
\end{figure}

\begin{figure}[tbp]
\setlength{\unitlength}{1mm}
\begin{center}
\includegraphics[width=80mm]{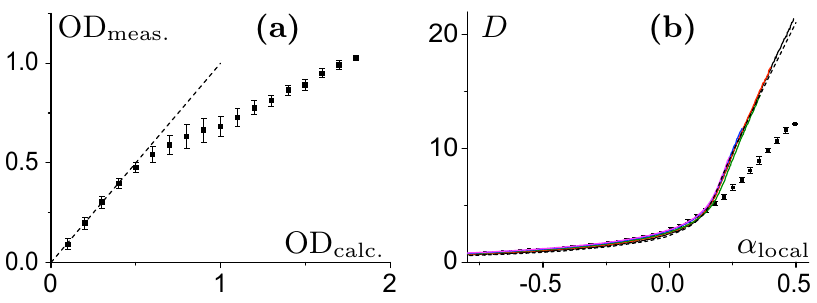}
\end{center}
\vskip -3mm
\caption{(Color online)  Combination of the experimental and theoretical results of Fig. \ref{fig:insitu}. Error bars indicate the standard deviation of the data. 
 (a) Measured OD as a function of calculated OD, averaged over the data shown in Fig. \ref{fig:insitu}. The dashed line with a slope 1 is a guide to the eye. (b) Continuous lines: QMC results for $D$ as a function of $\alpha_{\rm local}=\alpha-m\omega^2r^2/\kB T$ for the data shown in Fig.~\ref{fig:insitu} (same color code).  Black dashed line: prediction of \cite{Prokofev:2002} for the uniform case. Dots: Measured $D$, averaged over all experimental data shown in Fig. \ref{fig:insitu}. 
 }
   \label{fig:PSD}
\end{figure}

We now compare our experimental and numerical results, first confronting the measured in-situ optical densities with theoretical profiles calculated for the same $\mu$ and $T$. As illustrated in Fig.~\ref{fig:insitu} (a-b), the wings of the calculated and measured profiles nearly coincide. However there is a clear discrepancy in the central part of the density distributions. Whereas the central OD in the four coldest experimental distributions is $\sim 1.0$, the QMC simulations systematically predict a central OD $\sim 1.8$, i.e. a density of $\sim 55\;\microns^{-2}$. A global comparison between predicted and measured OD is shown in Fig.~\ref{fig:PSD}a, where we performed an average  over the 5 profiles of 
Fig.~\ref{fig:insitu}. 

We now discuss several possible causes for this discrepancy. A first possible source of error is the uncertainty on the  detectivity factor $\xi$. To estimate its influence, we have reprocessed the measured profile shown in Fig.~\ref{fig:insitu}a, by choosing the lower ($\xi=0.21$) and upper ($\xi=0.29$) values of the uncertainty interval for $\xi$.  The QMC results for the modified fit parameters are shown with dash-dotted and dashed lines in Fig.~\ref{fig:insitu}a. Clearly, the uncertainty on $\xi$ does not account for the observed deviation. Other `technical' causes for this discrepancy could be the imperfect resolution of the imaging system and/or the atomic motion during the imaging pulse. However, neither of them can account for the difference between predicted and measured density profiles \cite{optics}. The most probable cause of this discrepancy is the reduced absorption cross-section for large 2d atomic densities, due to multiple scattering of the photons of the probe laser beam. Although our optical densities ($\lesssim 1$) do not exceed usual values for absorption imaging, they correspond in this 2d geometry to a short mean distance $d$ between scatterers. For the densest clouds we find that $k_{\rm L}d$ is on the order of 1 ($k_{\rm L}=8\times 10^6\,$m$^{-1}$), which can significantly modify the photon scattering rate  \cite{Sokolov:2009,future}.

The trapped system of interest here can be related to a uniform gas using the local density approximation (LDA). Within LDA, the phase space density $D(\bs r)$ in a trapping potential $V(\bs r)$ is given by  $D(\bs r)=F(\mu-V(\bs r), T)$ where $D=F(\mu, T)$ is the equation of state of the uniform system. For the quasi-2d Bose gas, the validity of LDA has been accurately checked with QMC simulations in \cite{Holzmann:2009red}. To test the scale invariance of the equation of state, we have plotted in Fig.~\ref{fig:PSD}b the phase space density $D$ as a function of the ratio $\alpha_{\rm local}\equiv\alpha(r)$ between $\mu -m\omega^2r^2/2$ and $\kB T$, using the same data as in Fig.~\ref{fig:insitu}. The calculated functions $D(\alpha (r))$ (continuous lines) nicely superpose onto each other, confirming the scale invariance in this low temperature region where the excitation of the $z$ motion does not play an important role. Note that the central phase space densities notably exceed the critical value for superfluid transition ($D_{\rm c}\simeq 8.0$ for $\tilde g=0.146$ \cite{Prokofev:2001}), signaling the presence of a significant superfluid component. The QMC results confirm the prediction of \cite{Prokofev:2002} for the uniform 2d gas, which was obtained using a classical field Monte Carlo method (dashed line in  Fig.~\ref{fig:PSD}b). The small corrections to the results of [8] are due to two factors: (i) the presence of residual excitations in the z-axis, and (ii) the finite value of the interaction parameter $\tilde g$ \cite{Holzmann:2009red}. Obviously the deviation between experimental and numerical data that was appearing in the density profiles of Fig.~\ref{fig:insitu} also shows up in the plot of $D(\alpha(r))$ in Fig.~\ref{fig:PSD}b, and the experimental values of $D$ (dots) lie systematically below the predicted ones in the high phase space region.

A simple way to circumvent the problem of imaging high density regions is to take advantage of another known scale invariance of the 2d Bose gas. This invariance manifests itself in a 2d ballistic expansion after the confinement in the $xy$ plane has been suddenly released. It follows from a hidden SO(2,1) symmetry of the 2d Bose gas  with contact interactions $V(\bs r)=(\hbar^2 \tilde g/m) \delta(\bs r)$ when it is confined in an isotropic harmonic potential of frequency $\omega$ \cite{Pitaevskii:1997a}. Starting from an arbitrary initial equilibrium profile $n_{\rm eq}(\bs r)$, the density profile after a TOF duration $t$ is obtained by a scaling transform
\begin{equation}
n(\bs r,t)=\eta_t^2\,n_{\rm eq}(\eta_t\bs r)\ , \qquad \eta_t=(1+\omega^2t^2)^{-1/2}\ .
\label{eq:scaling}
\end{equation}
This relation was predicted within the Bogoliubov approximation in \cite{Kagan:1996b} and it holds exactly for interaction potentials that satisfy $V(\lambda \bs r)=V(\bs r)/\lambda^2$ \cite{Pitaevskii:1997a}. The 2d contact potential belongs to this class of functions, although strictly speaking it needs to be regularized at short distances to avoid ultraviolet divergences.

\begin{figure}[tbp]
\begin{center}
\includegraphics[width=80mm]{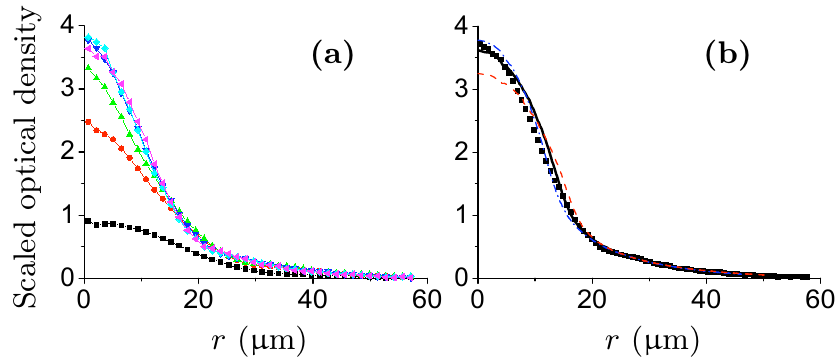}
\end{center}
\vskip -3mm
\caption{ (Color online) (a) Optical density profiles obtained for a TOF duration $t=0$ (black), 3 (red), 6 (green), 10 (blue), 12 (cyan), 14 (magenta) ms and rescaled to their in-situ value according to (\ref{eq:scaling}). (b) Squares: Optical density profile obtained by averaging the results of (a) for $10\leq t \leq 14$\,ms, yielding fit parameters $T=94$\,nK, $\alpha=0.36$ for $\xi=0.63$. Lines: QMC results for the same fit parameters (continuous, $N=42000$), and for those deduced assuming $\xi=0.47$ [dashed red, ($T\; {\rm(nK)},\alpha,N$)= (104,\,0.39,\,57600)] and $\xi=0.79$ [dash-dotted blue, (87,\,0.33,\, 32100)].}
  \label{fig:longTOF}
\end{figure}

Experimentally, we initiate the 2d expansion by switching off the magnetic trap while keeping the optical potential constant. The atom cloud thus expands in the $xy$ plane for an adjustable duration $t$ after which we take an image of the cloud.  We explore TOF durations up to $t=14$~ms for which  the central density is reduced by a factor $\eta_t^{-2}=4$, so that artifacts due to multiple scattering of probe photons should be strongly reduced (see Fig.~\ref{fig:PSD}a). We show in Fig.~\ref{fig:longTOF}a a succession of density profiles recorded for TOF durations varying from 0 to 14 ms. Each profile has been rescaled to the initial in-situ distribution according to the law (\ref{eq:scaling}), so that ideally all profiles should be superimposed. In practice this superposition is poor for short TOF durations and becomes better as the clouds expand. All scaled profiles obtained for $t\geq 10\,$ms coincide within their noise, showing that the detectivity is constant. We find $\xi=0.63\,(16)$, consistent with the measured absorption line-width of the probe laser. Moreover, the agreement between these scaled profiles and the calculated ones is indeed very satisfactory (see Fig.~\ref{fig:longTOF}b)\cite{anisotropy}. Note that the scale invariance accessible with such 2d expansions is notably different from the properties revealed in a 3d expansion of an initially quasi-2d gas, for which all trapping potentials are switched off simultaneously. In that case interactions are negligible during the TOF, due to the fast expansion along the initially strongly confined direction. This gives access to the quasi-coherent core which is present at the center of the trap below the Berezinskii--Kosterlitz--Thouless transition \cite{Hadzibabic:2006,Kruger:2007,Clade:2009}. 

In conclusion we have studied the equilibrium density profile of a trapped quasi-2d Bose gas. The discrepancy between in-situ absorption images and numerical calculations suggests that multiple scattering of the probe photons reduces the absorption cross-section in high-density regions \cite{future}.  To reveal the undistorted density distribution of the gas, we have taken advantage of its self-similarity in a 2d expansion. The profiles measured after this expansion are in good agreement with the QMC predictions. Natural extensions of this work are the measurement of other thermodynamical quantities from in-situ images using the procedure proposed in \cite{Ho:2009}, and the study of the regime where the scale invariance breaks down and energy-dependent corrections to $\tilde g$ become important \cite{Schick:1971,Popov:1972,Lim:2009,Mora:2009,astrakharchik:2009red}.


\begin{acknowledgments}
K.~J.~G. and S.~P.~R. acknowledge support from the EU (contract IEF/219636) and from the German Academic Exchange Service (DAAD, grant D/06/41156), respectively. This work is supported by IFRAF, ANR (project BOFL) and SCALA. 
\end{acknowledgments}


\begin{thebibliography}{24}
\bibitem{optics}
The modelling of our lens triplet imaging system gives a half-width at half maximum value of $\sim 2\,\microns$ for the point-spread response function, much smaller than the typical radius of the gas ($20\,\microns$). The transverse displacement of the atoms during the imaging pulse (duration $60\,\mu$s) can be accounted for by a convolution of the initial density profile with a gaussian of width $\lesssim 1.5\,\microns$. Also, their longitudinal displacement $\lesssim 30\,\microns$ is too small to significantly defocus the image.
\bibitem{future}
A detailed modeling of atom-light interaction in this dense quasi-2d regime will be presented elsewhere.
\bibitem{anisotropy}
Small deviations to the scaling (\ref{eq:scaling}) may be due to the slight anisotropy of the trap in the $xy$ plane.
\end{thebibliography}

\end{document}